\documentclass[graybox]{svmult}

\usepackage{mathptmx}       
\usepackage{helvet}         
\usepackage{courier}        
\usepackage{type1cm}        
\usepackage{makeidx}         
\usepackage{graphicx}        
\usepackage{multicol}        
\usepackage[bottom]{footmisc}

\usepackage{url}             
\usepackage{xcolor}          
\usepackage{hyperref}        
\usepackage{braket}          

\usepackage{dsfont}          
\usepackage[caption=false]{subfig}       
\usepackage{placeins}        

\DeclareMathAlphabet{\mathcal}{OMS}{cmsy}{m}{n}
\usepackage{mathtools}

\makeindex             

\begin{document}

\title*{An Introduction to Superconducting Qubits and Circuit Quantum
Electrodynamics}
\titlerunning{Introduction to Superconducting Qubits}
\author{Nicholas Materise}
\institute{Nicholas Materise \at Lawrence Livermore National Laboratory\\ 7000 East
Ave\\ Livermore, CA 94550\\ \email{materise1@llnl.gov}}

\maketitle

\vspace{-30mm}

\abstract*{Superconducting qubits have matured from platforms demonstrating and manipulating macroscopic coherent quantum states to realizing exotic quantum states, running surface error correction codes, and single photon detection to name a few recent milestones. This article will review the fundamentals of circuit QED related to the design and simulation of superconducting qubits. }

\section{Introduction}
\label{sec:intro}
Superconducting qubits and circuit quantum electrodynamics have enabled design
of solid state sources of quantum information. The performance of these devices
has scaled exponentially over the last fifteen years, in terms of their energy
relaxation and dephasing times, drawing interest from adjacent communities
including the Axion Dark Matter Experiment (ADMX). Recently, superconducting
qubits have been targeted for use as single photon detectors in the ADMX high
frequency experiment, ADMX-HF~\cite{Shokair2014}. The goal of this article
is to give members of the ADMX community an introduction to some of the models
used to analyze and design superconducting qubits. This review is not an
exhaustive coverage of the field, but it aims to guide the reader to relevant
literature and analysis techniques that closely follow experiment.

\section{Superconducting Qubit Circuit Models} \label{sec:circuit_models}

A qubit is a two level system or a system whose controllable quantum dynamics
involve its two lowest lying energy levels. Nature provides several forms of qubits
or carriers of quantum information including single photons, trapped ions, and
atoms in high finesse cavities. Superconducting qubits realize
\textit{artificial atoms} with engineered energy levels using the non-linearity of
Josephson junctions and surrounding microwave circuitry~\cite{Blais2004}. 
The quantum dynamics of these systems follows
that of a damped and driven anharmonic oscillator whose 
anharmonicity is controlled by choice
of circuit parameters, e.g. linear capacitance and inductance of the
Josephson junction~\cite{Schuster2007}. For experimental design and control, 
practitioners draw from the Jaynes-Cummings model and
its variants from cavity quantum electrodynamics (QED)~\cite{Jaynes1963,Schuster2007}.
Circuit quantum electrodynamics borrows the application of second quantized
Hamiltonians from atomic optics via a standardized procedure for quantizing
passive circuit. This section will introduce simple models for Josephson
junctions and their role in superconducting qubits. We will then discuss circuit
quantization methods and Black box quantization techniques used to obtain
second quantized Hamiltonians.

\subsection{Non-linearity in Superconducting Qubits}
\label{subsec:jj_nonlin_qubits}
\begin{figure}[ht]
\centering
\subfloat[]{%
    \includegraphics[%
    width=0.45\textwidth]%
    {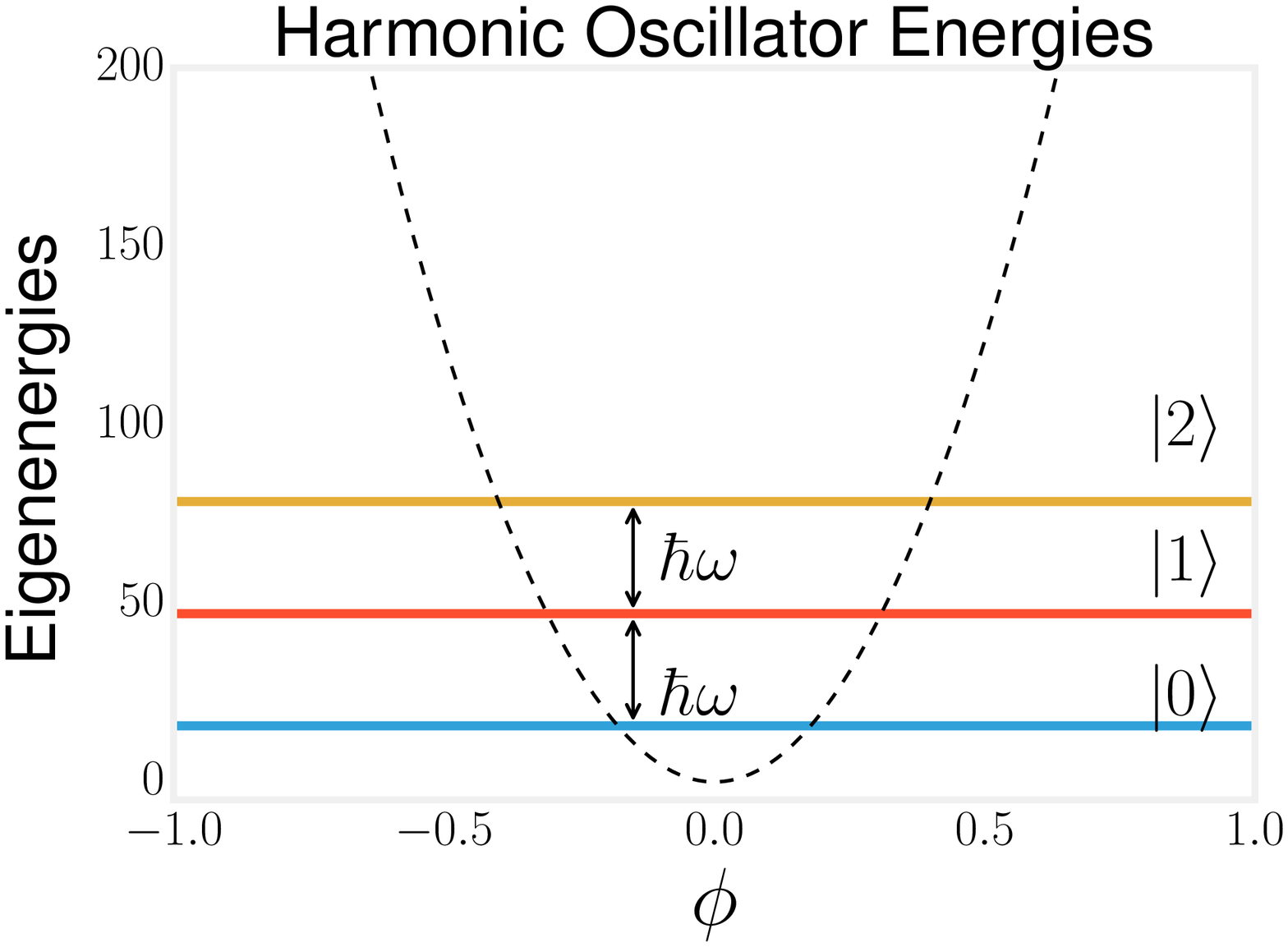}}%
\qquad
\subfloat[]{%
    \includegraphics[%
    width=0.45\textwidth]%
    {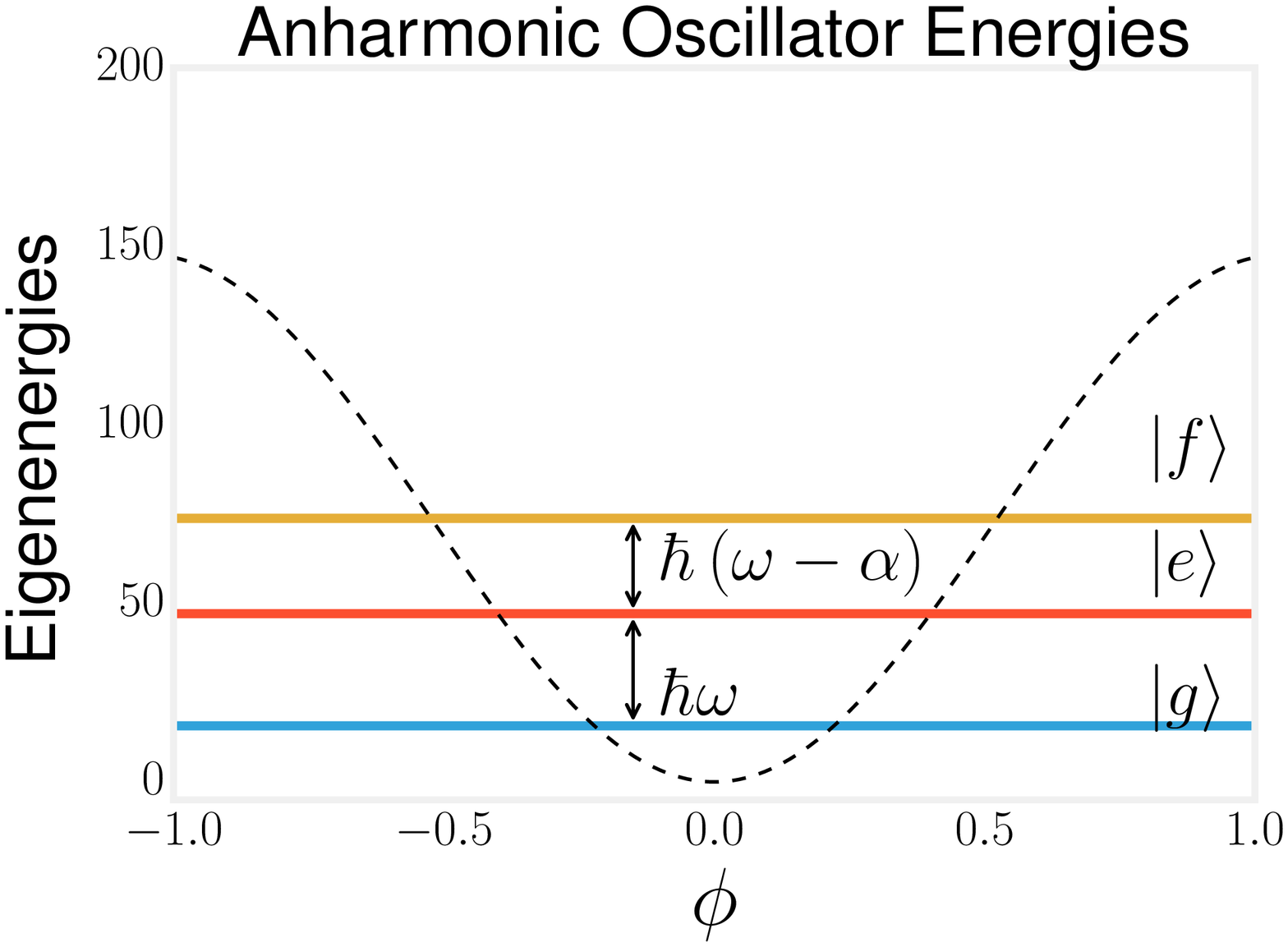}} \\
\subfloat[]{%
    \includegraphics[%
    width=0.18\textwidth]%
    {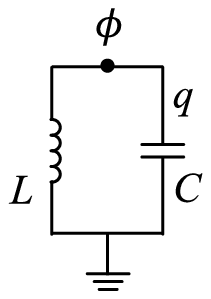}}%
\hspace{32mm}
\subfloat[]{%
    \includegraphics[%
    width=0.22\textwidth]%
    {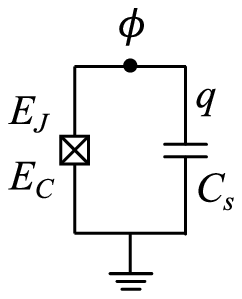}}%
\caption{Comparison of the quantum harmonic oscillator with anharmonic
oscillator. (a) and (b) give the eigenenergies of the two oscillators, where the
horizontal lines are the eigenenergies and the dashed lines represent notional
potentials. (c) and (d) are the corresponding circuit models for
 an LC circuit and a transmon qubit~\cite{Koch2007}.}
\label{fig:lc_circuit}
\end{figure}

The operational modes of superconducting qubits vary by their energy spectra,
where non-linearity plays a role in realizing accessible and isolated states.
If we consider the lowest two levels of the quantum harmonic oscillator to be
the ground and excited states of a qubit $\left(\ket{g},\ \ket{e}\right)$, 
 the energies for the two states are separated by integer multiples of
$\hbar\omega$. The classical electric circuit model for 
an oscillator is the LC circuit, shown
in Figure~\ref{fig:lc_circuit}. We will refer to this model in
Section~\ref{subsec:cqed} when we derive the second quantized form of the
Hamiltonian for an LC circuit. Figure~\ref{fig:lc_circuit} compares the LC
oscillator circuit to an anharmonic qubit, the transmon. Notice that the spacing
between the excited state $\ket{e}$ and the next highest state $\ket{f}$ is
smaller than the spacing between $\ket{g}$ and $\ket{e}$. In more anharmonic
oscillators, the spacing is larger, further isolating the qubit states from the other
states of the oscillator. The transmon trades off its anharmonicity for reduced
sensitivity to charge noise~\cite{Blais2004}.

\FloatBarrier
\FloatBarrier

An anharmonic oscillator-based superconducting qubit inherits its non-linearity
from Josephson junctions, where the non-linearity is tunable through fabrication
and microwave circuit design. To develop some intuition 
for the dynamics of Josephson junctions,
we will discuss classical circuit models for the device and 
their role in superconducting qubits.

\subsection{Classical Circuit Models of Josephson Junctions}
\label{subsec:jj_models}
There are several phenomenological models for Josephson junctions that are
motivated by the underlying device physics and limits of the electric circuit
analogs. We will review the Resistive and Capacitively Shunted Junction (RCSJ)
model as outlined in~\cite{Gross2016}.

\FloatBarrier
\begin{figure}[ht]
\centering
\subfloat[]{%
    \includegraphics[%
    width=0.125\textwidth]%
    {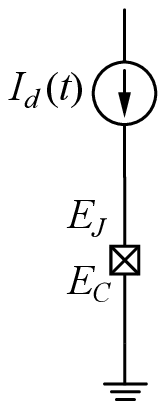}}
\qquad
\subfloat[]{%
    \includegraphics[%
    width=0.2\textwidth]%
    {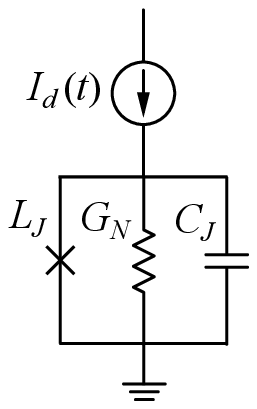}}
\qquad
\subfloat[]{%
    \includegraphics[%
    width=0.25\textwidth]%
    {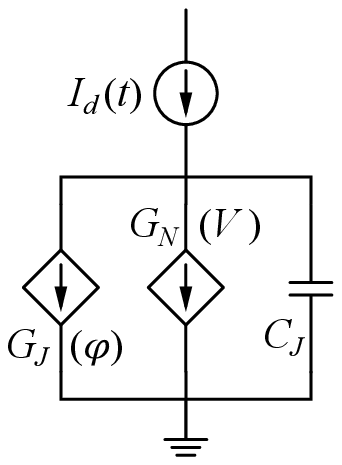}}
\caption{(a) Circuit diagram for a current driven Josephson junction,
(b) RCSJ circuit model, (c) equivalent circuit with current sources replacing
the conductance $G_N$ and inductor $L_J$}
\label{fig:rcsj_circuit}
\end{figure}
\FloatBarrier

In Figure~\ref{fig:rcsj_circuit} above, the left most circuit shows a
current-driven Josephson with drive current, $I_d$. The junction is approximated
as the parallel combination of an inductor $L_J$, conductance $G_N$, and
capacitor $C_J$. We replace the inductor and conductance with two voltage
controlled current sources (VCCS's), $G_J(\varphi), G_N(V)$, where we use the
\texttt{Gxxx} VCCS notation from SPICE~\cite{Nagel1973}. Kirchhoff's current law
at the node joining the three circuit elements with the drive current source
reads~\cite{Gross2016}

\begin{eqnarray}
I_d(t) = I_c\sin\varphi + V G_N(V) + C_J \frac{dV}{dt} \label{eq:Id} \\
G_N(V) = \left\{
\begin{array}{cc} 
0, &  \left|V\right|\leq 2\Delta_0/e\\
1/R_N, & \left|V\right| \geq 2\Delta_0/e
\end{array}
\right.\label{eq:GN}
\end{eqnarray}

All occurrences of $V$ refer to the voltage across the three elements
representing the Josephson junction from the node of their intersection to
ground. The superconducting gap energy at zero temperature, $\Delta_0$, gives
the voltage where the junction transitions from superconducting to a normal
metal with a normal resistance $R_N$, see Eq.~\ref{eq:GN}. For finite
temperatures,(Gross et al. 2016) gives the temperature dependent conductance 
in the RCSJ based on the the density of states of
quasiparticles in the Josephson junction~\cite{Gross2016}.

The VCCS $G_J(\varphi)$ varies sinusoidally with the junction phase,
$\varphi$, which is a function of the voltage across the junction and given
by the Josephson equation

\begin{eqnarray}
V = \frac{\Phi_0}{2\pi}\frac{d\varphi}{dt} \label{eq:jj_phi_v} \\
\Phi_0 = \frac{h}{2e} \equiv \mathrm{Magnetic\ flux\ quantum} \nonumber
\end{eqnarray}

If we substitute Eq.~\ref{eq:jj_phi_v} into Eq.~\ref{eq:Id}, we arrive at a
second order linear differential equation in the phase variable, $\varphi$

\begin{equation}
I_d(t) = \frac{\Phi_0}{2\pi}C_J\frac{d^2\varphi}{dt^2} %
+\frac{\Phi_0}{2\pi}\frac{d\varphi}{dt} %
G_N\left(\frac{\Phi_0}{2\pi}\frac{d\varphi}{dt}\right) %
+ I_c\sin\varphi\label{eq:phi_ode2}
\end{equation}

This equation is analogous to a driven pendulum, where the capacitance and
conductance are proportional to the mass and damping parameter for the pendulum,
respectively~\cite{Schuster2007}. For practical, classical simulations of
Josephson junctions, the two VCCS model shunted by the junction capacitance
is sufficient to produce hysteresis in the current-voltage (IV) characteristic
curve. Numerical simulation of the circuit in Figure~\ref{fig:rcsj_circuit} is
well suited for SPICE~\cite{Nagel1973} circuit solvers or coupled to
geometries in multiphysics codes such as COMSOL~\cite{COMSOL}.

The RCSJ model is an intuitive model for the behavior of a Josephson junction
with an applied dc or ac drive current, though it is not as suitable for
superconducting qubit design and simulation. Circuit Quantum Electrodynamics
provides a framework analyzing such systems with the language of atomic optics
or cavity quantum electrodynamics. We will examine the key features of circuit
QED and its utility in the design and simulation of superconducting qubits.

\subsection{Circuit Quantum Electrodynamics} \label{subsec:cqed}
Circuit quantum electrodynamics (QED) combines microwave engineering, circuit
analysis, and quantum optics. Fabry-Perot cavities from optics are replaced by
resonant microwave cavities or lumped element microwave resonators in circuit
QED. The procedure for obtaining the quantized Hamiltonian and subsequent
dynamics of the system follows first from a classical treatment, then
quantization of the classical variables as operators and relating those
operators to bosonic single-mode raising and lowering operators
$\left\{\hat{a}_i^{(\dagger)}\right\}$.

\subsubsection{Quantizing the LC Oscillator} \label{subsubsec:lc_oscillator}
We return to the LC oscillator circuit in Figure~\ref{fig:lc_circuit} and write
the Lagrangian for the circuit in terms of the flux variable $\phi$ which is
treated as the generalized coordinate for the system~\cite{Clerk2010}.

\begin{equation}
\mathcal{L}\left(\phi, \dot{\phi}\right) = \frac{1}{2}C\dot{\phi}^2 %
- \frac{1}{2L}\phi^2\label{eq:lc_lagrangian}
\end{equation}

We treat the charge $q$ on the capacitor as the conjugate momentum and perform a Legendre
transformation to obtain the Hamiltonian as a function of both $q$ and $\phi$.

\begin{eqnarray}
q = \frac{\partial\mathcal{L}}{\partial\dot{\phi}} = C\dot{\phi} %
\Longrightarrow \dot{\phi}^2 = q^2 / C^2 \nonumber \\
\mathcal{H}\left(q, \phi\right) = \dot{\phi} q - \mathcal{L}%
= \frac{1}{2} C \dot{\phi}^2 + \frac{1}{2L}\phi^2 \nonumber \\
\mathcal{H} = \frac{1}{2C} q^2 + \frac{1}{2L}\phi^2 \label{eq:lc_H_classical}
\end{eqnarray}

Following the example in Chapter 3 of~\cite{Schuster2007}, the charge and flux
variables are quantized by converting them to operators with the commutation
relation $\left[\hat{\phi}, \hat{q}\right]=i\hbar$. If we take the resonance
frequency of the LC circuit to be $\omega=\left(LC\right)^{-1/2}$ and replace
$1/L$ in the potential term of the Hamiltonian, we arrive at the familiar form
for a harmonic oscillator with mass $C$.

\begin{equation}
\mathcal{H}\to\hat{H} = \frac{\hat{q}^2}{2C} %
+ \frac{1}{2}C\omega^2\hat{\phi}^2
\end{equation}

We define raising and lowering operators for this quantum harmonic
oscillator in analogy to those used in the one-dimensional model
and write the second quantized form of the Hamiltonian.

\begin{eqnarray}
\hat{q} = -i\sqrt{\frac{\hbar\omega C}{2}} %
\left(\hat{a} - \hat{a}^{\dagger}\right), %
\ \ \ \ \hat{\phi} = \sqrt{\frac{\hbar}{2\omega C}} %
\left(\hat{a} + \hat{a}^{\dagger}\right) \label{eq:q_phi_ops} \\
\hat{H} = \hbar\omega\left(\hat{a}^{\dagger}\hat{a} + 1/2\right)
\label{eq:lc_H_quantum}
\end{eqnarray}

\subsubsection{Black Box Circuit Quantization}
\label{subsubsec:bbq}
In the previous section, we covered a procedure for quantizing an LC oscillator
circuit which leads to an approximate generalization for any device given its
frequency dependent impedance function. This approach connects full wave
electromagnetic simulations of microwave circuits to their quantum mechanical
analogs in circuit QED. Given a single port $S$-parameter as a function of
frequency, one can obtain the impedance at the port by the transformation

\begin{eqnarray} 
Z = \left(\mathds{1} + S\right)\left(\mathds{1} - S\right)^{-1}\label{eq:s2z} \\
\mathds{1}\equiv\mathrm{identity\ matrix\ with\ same\ dimensions\ as}\ S \nonumber
\end{eqnarray}

Following the \textit{Black box quantization} methods outlined
in~\cite{Nigg2012, Solgun2014}, the impedance function, $Z(\omega)$ can be
expressed as a pole-residue expansion in the complex frequency $s=j\omega$,
where $j=-\sqrt{-1}$, following the electrical engineering convention.

\begin{eqnarray}
Z(s) = \sum_{k=1}^{M}\frac{r_k}{s - s_k} + d + es\label{eq:z_vfit}\\
\left\{r_k = a_k + jb_k\right\} \equiv \mathrm{residues}, \ \ \ \ %
\left\{s_k = \xi_k + j\omega_k\right\} \equiv \mathrm{poles} \nonumber
\end{eqnarray}

The above rational function can be obtained by a least squares fit of the
original impedance using the Vector Fit software outlined
in~\cite{Gustavsen1999} and available at~\cite{VectorFit}. If we take the case
where $d=0$ and the pole at $s\to\infty$ vanishes or $e=0$ and perform the
following partial fraction expansion and approximation for the $k$-th term in
the series and we find the $k$-th term is the impedance for a parallel RLC
oscillator circuit.

\begin{eqnarray}
Z_k(s) = \frac{r_k}{s - s_k} = \frac{r_k}{s - s_k} + \frac{r_k^*}{s - s_k^*}%
\simeq \frac{2a_k s}{s^2 - 2\xi_k s + \omega_k^2}\nonumber\\
\Longrightarrow Z_k(s) = \frac{\frac{\omega_kr_k}{Q_k}s}%
{s^2 + \frac{\omega_k}{Q_k}s + \omega_k^2}\label{eq:z_bb_approx} \\
\omega_k = \left(L_kC_k\right)^{-1/2}, \ \ \ \ %
Q_k = \omega_kR_kC_k = -\omega_k/2\xi_k, \ \ \ \ R_k = -a_k/\xi_k \nonumber
\end{eqnarray}

The total impedance, $Z(s)$ is a series combination of RLC oscillators and if we
take the dissipationless limit by ignoring the resistances, we can treat $Z(s)$
as a series combination of LC circuits and apply the same analysis from
Section~\ref{subsubsec:lc_oscillator} to each subcircuit.
If we shunt the resulting circuit, with a single Josephson junction, we can
obtain a simple model for the Hamiltonian of a qubit coupled
to a superconducting resonator with $M$-modes. For a full derivation of the
non-linear components of the Hamiltonian $\hat{H}_{\mathrm{nl}}$, see~\cite{Nigg2012};
we reproduce the salient features here.

\begin{eqnarray} 
\hat{H} = \hat{H}_0 + \hat{H}_{\mathrm{nl}} \\
\hat{H}_0 = \sum_i\hbar\omega_i\hat{a}_i^{\dagger}\hat{a}_i, \ \ \ \ %
\hat{H}_{\mathrm{nl}} = E_J\left(1 - \cos\hat{\varphi}
-\frac{\hat{\varphi}^2}{2}\right) \\ %
\hat{H}_{\mathrm{nl}}\approx -\frac{1}{2}\sum_i\alpha_i%
\hat{a}_i^{\dagger 2}\hat{a}_i^2%
-\sum_{i\neq j}\chi_{ij}\hat{a}_i^{\dagger}\hat{a}_i%
\hat{a}_j^{\dagger}\hat{a}_j \label{eq:Hnl_dispersive}\\
\hat{\varphi} = \frac{2\pi}{\Phi_0}\sum_i\hat{\phi}_i%
=\frac{2\pi}{\Phi_0}\sum_i\sqrt{\frac{\hbar}{2\omega_iC_i}}%
\left(\hat{a}_i + \hat{a}_i^{\dagger}\right)
\end{eqnarray}

The Hamiltonian above is referred to as the dispersive Hamiltonian for a weakly
anharmonic qubit coupled to a series of harmonic modes. In the
non-linear term, $\hat{H}_{\mathrm{nl}}$, the first contribution describes the
anharmonicities of those modes and the qubit mode or self-Kerr terms and the second
term gives the cross-Kerr terms~\cite{Nigg2012}. Both $\{\alpha_i\}$ and
$\{\chi_{ij}\}$ are experimentally observable, tying this model for
qubit-circuit interactions to physical devices.

\FloatBarrier
\begin{figure}[ht]
\centering
\includegraphics[width=0.8\textwidth]{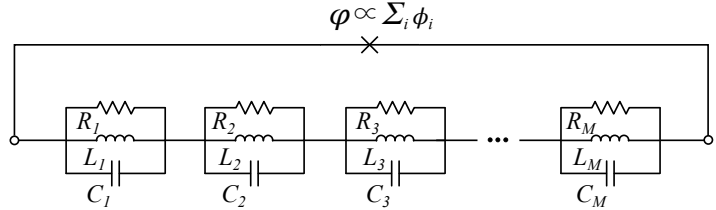}
\caption{Series combination of RLC circuits shunted by a single Josephson
junction representing the black box circuit from Eq.~\ref{eq:z_bb_approx} and
similar in design to the circuit in~\cite{Bourassa2012}}
\label{fig:bbq_lc_array}
\end{figure}
\FloatBarrier

\section{Summary} \label{sec:summary}
The models used to describe the operation of superconducting qubits follow
intuitive modifications to the familiar damped and driven oscillator systems
from classical and quantum mechanics. These models arise from careful
application of circuit QED to incorporate the quantum effects of macroscopic
structures in microwave circuits. Although the dispersive Hamiltonian describes
many superconducting qubit systems in quantum information experiments, this
article did not apply the model to the problem single photon counting. For
more resources related to circuit QED and single photon counting, please refer
to~\cite{Clerk2010,Schuster2007,Bishop2010,Johnson2011,Reed2013,Holland2015}.

\section{Acknowledgements}
This work was performed under the auspices of the U.S. Department of Energy by
Lawrence Livermore National Laboratory under Contract DE-AC52-07NA27344 and
funded by the Laboratory Directed Research and Development programs at LLNL
project numbers 15-ERD-051, 16-SI-004.

\pagebreak
\bibliographystyle{abbrv}

\end{document}